# A FRACTAL APPROACH TO CHARACTERIZE EMOTIONS IN AUDIO AND VISUAL DOMAIN: A STUDY ON CROSS-MODAL INTERACTION

Sayan Nag[a,e], Uddalok Sarkar[a,e], Shankha Sanyal[a,f]*, Archi Banerjee[a,b,c], Souparno Roy[a,d], Samir Karmakar[f], Ranjan Sengupta[a] & Dipak Ghosh[a]

[a]Sir C.V. Raman Centre for Physics and Music, Jadavpur University
[b] Department of Humanities and Social Sciences, IIT Kharagpur
[c]Shrutinandan School of Music, Kolkata
[d]Department of Physics, Jadavpur University
[e]Department of Electrical Engineering, Jadavpur University
[f]School of Languages and Linguistics, Jadavpur University

* Corresponding Author email: ssanyal.sanyal2@yahoo.com

**Abstract:**
*It is already known that both auditory and visual stimulus is able to convey emotions in human mind to different extent. The strength or intensity of the emotional arousal vary depending on the type of stimulus chosen. In this study, we try to investigate the emotional arousal in a cross-modal scenario involving both auditory and visual stimulus while studying their source characteristics. A robust fractal analytic technique called Detrended Fluctuation Analysis (DFA) and its 2D analogue has been used to characterize three (3) standardized audio and video signals quantifying their scaling exponent corresponding to positive and negative valence. It was found that there is significant difference in scaling exponents corresponding to the two different modalities. Detrended Cross Correlation Analysis (DCCA) has also been applied to decipher degree of cross-correlation among the individual audio and visual stimulus. This is the first of its kind study which proposes a novel algorithm with which emotional arousal can be classified in cross-modal scenario using only the source audio and visual signals while also attempting a correlation between them.*
**Keywords: Cross-modal valence, Emotions, Audio/visual stimuli, 2D-DFA, Hurst Exponent**

**INTRODUCTION**
A number of studies have been done in the psychological level as to how the emotions conveyed by two different modalities - audio and visual vary from one another. A few of them even try to look into how the perceptual strength of emotions expressed in the two modalities differs from one another. It is already known that certain emotions (i.e., brief affective states triggered by the appraisal of an event in relation to current goals [4]) such as awe and wonder [5] are frequently reported in relation to the contemplation of artworks. These emotions typically occur when an object or event is appraised as highly complex and novel and creates a sense of being in the presence of something greater than oneself [6]. However, it has also been recently emphasized that affective responses to art are more diverse, and often include emotions such as sadness [7] and nostalgia [8], which are also experienced in other everyday situations that do not involve contemplation of artworks. If we try to look into more basic features of a painting i.e. the usage of basic colors (Red, Green and Blue) along with their offshoots, earlier works suggest that warm colors – such as red, yellow and orange – can spark a variety of emotions ranging from comfort and warmth to hostility and anger. Cool colors – such as green, blue and purple – often spark feelings of calmness as well as sadness. But, all these are psychological studies based on human response data, till date there have been no study which computationally classifies the emotional appraisal corresponding to a group of paintings. In this work we have tried to evaluate long range temporal correlations corresponding to these three color components in paintings. In recent years, the use of musical stimuli as an important means of emotional appraisal is being developed with special focus on cross-modal transfer of emotions. While most of these studies look into the psychological and cognitive aspects of the musical and visual stimulus, the source characteristics of these stimuli are largely neglected mainly due to the lack of robust features to quantify them. The development of the International Affective Picture System (IAPS) has been followed by a similar collection of sounds, the International Affective Digitized Sounds (IADS) – a series of naturally occurring human, non-human, animal, and environmental sounds (e.g., bees buzzing; applause, explosions). In two experiments by Bradley and Lang (2000), it was shown that valence and arousal ratings of these sounds were comparable to affective pictures from the IAPS. On a physiological level, emotionally arousing sounds elicit large electrodermal activity, which is generally known to be sensitive to the arousal of emotional stimuli.

In this paper, for the first time, we look to classify the emotional sound and visual stimuli solely from their source characteristics, i.e. the time series generated from the audio signal and the 2-dimensional matrix of pixels generated from the affective picture stimulus. The sample data consists of 6 audio signals of around 10 second each and 6 affective pictures, of which 3 each belonged to positive and negative valence respectively. The emotional ratings corresponding to the visual and audio stimulus were standardized a-priori with the help of different psychological tests and corroborated with standardized measures present in literature. The main aim of this work is to ratify the results of psychological tests in the perceptual domain with the mathematical quantitative output obtained from the source signals itself. As a powerful mathematical tool, fractal theory initiated by Mandelbrot in the 1960s has been widely applied to many areas of natural sciences. Since the simple iterative algorithm in the fractal theory can generate a variety of complex images, fractal dimension is considered as an effective measure of the complexity of the target object. We use a robust non-linear data analysis tool called Detrended Fluctuation Analysis (DFA) to calculate the long-range temporal correlations or the Hurst exponent corresponding to the auditory signals. Similar non-linear analyses have been previously used in the scientific community to comprehend the underlying complexities in these inherently convoluted audio signals [10 - 21]. On the other hand, the 2D analogue of the same DFA technique has been applied on the array of pixels corresponding to affective pictures of contrast emotions, which essentially gives the long-range spatial correlations of individual color components. We used the scaling exponent (or the Hurst exponent) obtained from the audio clips and the visual images as a robust parameter to quantify their emotional valence. Thus we have a single unique scaling exponent corresponding to the each 1D audio signal and three scaling exponents corresponding to RED/GREEN/BLUE (RGB) component in each of the visual images. In this way we have been able to provide a quantitative classification of emotional cues in auditory and visual domain using the source signals itself. Further, the correlation features among the paintings as well as the audio clips have also been computed using the 2D/1D- Detrended Cross-correlation (DCCA) technique, which essentially gives the degree of correlation between the individual domains. To conclude, for the first time we propose a novel algorithm with which emotional arousal can be classified in cross-modal scenario using only the source audio and visual signals while also attempting a correlation between them. The study is expected to go a long way in research of multimodal interaction of emotional cues in multiple domains. The results and implications have been discussed in detail.

**EXPERIMENTAL DETAILS:**
**Choice of three pairs of** *audio and visual stimuli*
6 clips of around 10 s each (3 clips each belonging to the positive and negative valence) were chosen from the IADS [1] database of musical clips and normalized for acoustic analysis. In a similar manner, 6 famous paintings were chosen of which 3 belonged to the positive and negative valence each. All the clips and paintings chosen were subjected to human response analysis of 50 participants to validate the emotional appraisal of each of the stimulus chosen. The details of the paintings and sound clips along with their mean ratings generated from the psychological response test prescribed by [9] are given below:

**Table 1: Psychological ratings of the audio clips chosen for analysis**

| Clip No. | Anger | Fear | Happy | Sad | **Target** |
|---|---|---|---|---|---|
| 1 | 1.00 | 1.00 | 7.33 | 1.00 | HAPPY |
| 2 | 1.00 | 1.00 | 7.17 | 1.17 | HAPPY |
| 3 | 1.00 | 1.00 | 7.17 | 1.00 | HAPPY |
| 4 | 1.17 | 1.00 | 1.00 | 7.67 | SAD |

| | 5 | 1.00 | 1.33 | 1.17 | 7.50 | SAD |
| | 6 | 1.00 | 1.67 | 1.00 | 7.50 | SAD |

**Table 2: Psychological ratings of the paintings chosen for analysis**

| Image No. (Painting Name) | Anger | Fear | Happy | Sad | Target | Painter |
|---|---|---|---|---|---|---|
| 1 (Sunflower) | 0.35 | 0.20 | 8.91 | 0.62 | HAPPY | Vincent Van Gogh |
| 2 (Japanese Vase) | 0.10 | 0.05 | 9.17 | 0.13 | HAPPY | Vincent Van Gogh |
| 3 (Almond Tree) | 2.12 | 0.25 | 8.29 | 0.23 | HAPPY | Vincent Van Gogh |
| 4 (The Tragedy) | 1.71 | 1.33 | 0.65 | 6.95 | SAD | Pablo Picasso |
| 5 (Starry Night) | 0.09 | 1.27 | 2.15 | 6.88 | SAD | Vincent Van Gogh |
| 6 (Sailboats at Sunset) | 0.90 | 0.32 | 2.75 | 6.35 | SAD | Ferdinand du Puigaudeau |

In this way, we have a standardized measure of each of the stimulus corresponding to both auditory and visual domain. A correlation study is performed across the cross-modal domain to establish the degree of emotional appraisal corresponding to the stimulus used.

## METHODOLOGY:

1-dimensional Detrended Fluctuation Analysis (DFA) is conventionally done following the algorithm of [2]. In this work, for extracting the scaling exponent corresponding to different paintings, we propose a novel 2D-DFA algorithm here:

### 2D-Detrended Fluctuation Analysis

This section describes the steps for computing Hurst Exponent using the two-dimensional DFA algorithm for a grayscale image $I$. The steps are as follows:

1) The profile $x_{i,j}$ is computed using:

$$x_{i,j} = \sum_{n=1}^{i} \sum_{m=1}^{j} (I_{i,j} - \bar{I})$$

where $m = 1, 2, \cdots, M$, $n = 1, 2, \cdots, N$, $I_{n,m} = 0, 1, \cdots, 255$ is the brightness of the pixel at the coordinates $(m, n)$ of the gray scale image and $\bar{I}$ represents the mean value of $I_{n,m}$.

2) $x_{i,j}$ is divided into small regions of size $s \times s$, where $s$ is set as:
$s_{min} \approx 5 \leq s \leq s_{max} \approx \min\{M, N\}/4$.

3) An interpolating curve is computed of $x_{i,j}$ using:

$$G_{i,j}(l, s) = a_l i + b_l j + c_l$$

in the $l^{th}$ small square region of size $s \times s$, which can be given by using a multiple regression procedure.

4) The variance in the $l^{th}$ small square region is computed for $s = s_{min}, s_{min} + 1, \cdots, s_{max}$, which is given by:

$$F_{i,j}^2(l, s) = \frac{1}{s^2} \sum_{n=1}^{i+s} \sum_{m=1}^{j+s} (x_{i,j} - G_{i,j}(l, s))^2$$

5) The root mean square $F(s)$ is computed as:

$$F(s) = \left[\frac{1}{L_s} \sum_{l=1}^{L_s} F_{i,j}^2(l, s)\right]^{1/2}$$

where $L_s$ denotes the number of the small square regions of size $s \times s$.

6) If $x_{i,j}$ has a long-range power-law correlation characteristic, then the fluctuation function $F(s)$ is observed as follows:

$$F(s) \propto s^{\alpha}$$

where α is the two-dimensional scaling exponent, a self-affinity parameter representing the long-range power-low correlation characteristics of the surface.

For investigating power law cross-correlations between different simultaneously recorded time series in the presence of nonstationarity, 1D-Detrended Cross correlation Analysis (DCCA) [3] has been used in many cases. Here we generalize it in 2-Dimensional analogue to extract the degree of correlation present between different paintings.

## 2D- Detrended Cross correlation Analysis (DCCA)

This section describes the steps for computing Cross-Correlation Coefficient using the two-dimensional DCCA algorithm for two grayscale images $A$ and $B$. The steps are as follows:

1) The profiles $x_{i,j}$ and $y_{i,j}$ are computed using:

$$x_{i,j} = \sum_{n=1}^{i} \sum_{m=1}^{j} (A_{i,j} - \bar{A})$$

$$y_{i,j} = \sum_{n=1}^{i} \sum_{m=1}^{j} (B_{i,j} - \bar{B})$$

where $m = 1, 2, \cdots, M$, $n = 1, 2, \cdots, N$, $A_{n,m} = 0, 1, \cdots, 255$, $B_{n,m} = 0, 1, \cdots, 255$ are the brightness of the pixel at the coordinates $(m, n)$ of the gray scale images and $\bar{A}$ and $\bar{B}$ represents the mean value of $A_{n,m}$ and $B_{n,m}$ respectively.

2) Both $x_{i,j}$ and $y_{i,j}$ are individually divided into small regions of size $s \times s$, where $s$ is set as:

$$s_{min} \approx 5 \leq s \leq s_{max} \approx \min\{M, N\}/4.$$

3) Interpolating curves are computed of $x_{i,j}$ and $y_{i,j}$ using:

$$Gx_{i,j}(l, s) = ax_l i + bx_l j + cx_l$$
$$Gy_{i,j}(l, s) = ay_l i + by_l j + cy_l$$

in the $l^{th}$ small square region of size $s \times s$, which can be given by using a multiple regression procedure.

4) The variance in the $l^{th}$ small square region is computed for $s = s_{min}, s_{min} + 1, \cdots, s_{max}$, which is given by:

$$F_{i,j}^2(l, s) = \frac{1}{s^2} \sum_{n=1}^{i+s} \sum_{m=1}^{j+s} \left(x_{i,j} - Gx_{i,j}(l, s)\right) * \left(y_{i,j} - Gy_{i,j}(l, s)\right)$$

5) The root mean square $F(s)$ is computed as:

$$F(s) = \left[\frac{1}{L_s} \sum_{l=1}^{L_s} F_{i,j}^2(l, s)\right]^{1/2}$$

where $L_s$ denotes the number of the small square regions of size $s \times s$.

6) If the profiles are long-range power-law correlated, then the fluctuation function $F(s)$ is observed as follows:

$$F(s) \propto s^\lambda$$

where $\lambda$ is the two-dimensional scaling exponent. The relation between cross-correlation exponent, $\gamma_x$ and scaling exponent $\lambda$ can be shown as:

$$\gamma_x = 2 - 2 * \lambda$$

For uncorrelated data, cross-correlation exponent has a value 1 and the lower the value of cross-correlation exponent more correlated are the data.

## RESULTS AND DISCUSSION

In the first part of our work, DFA exponent was computed for the 6 audio clips and the 6 paintings that were put to analysis. In case of the paintings, $\alpha_{red}$, $\alpha_{green}$ and $\alpha_{blue}$ were computed corresponding to the Red, green and blue color component of the painting analyzed. In the following figures, the DFA exponent corresponding to each clip and visual stimuli have been plotted. **Fig. 1** shows the scaling exponents for the audio clips which have been classified apriori as happy and sad; while **Fig. 2** denotes the scaling exponent for the paintings.

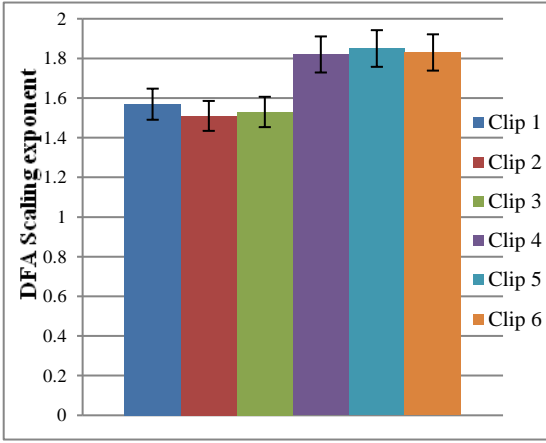 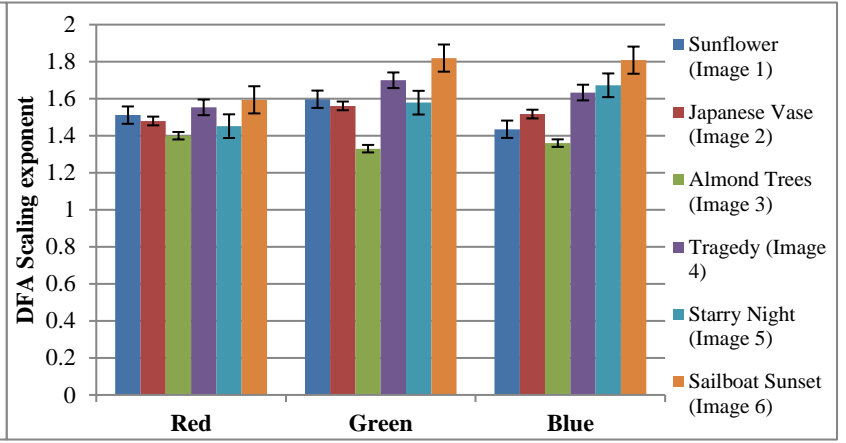

Fig. 1: DFA exponent of audio clips    Fig. 2: DFA exponents (color-wise) of visual stimuli

From **Fig. 1,** it is evident that the scaling exponents of Clips 1 to 3 are lower as compared to the scaling exponents of Clips 4 to 6 i.e. the LRTC present in Clips 1 to 3 are lower than the temporal correlations present in Clips 1 to 6. This can be attributed to various acoustic features of these clips like tempo, rhythm etc, but the mathematical manifestation is the decrease/increase in long range temporal correlations. In **Fig. 2**, again it is evident that the scaling exponents corresponding to Images 4 to 6 are in general higher than Images 1 to 3. Also, an interesting observation is that $α_{green}$ and $α_{blue}$, i.e. the scaling exponents corresponding to green and blue color show the maximum increase for Images 4 to 6 (which were classified as evoking sad emotions). Thus, the manifestation of sad emotion can be attributed to higher order of correlations present in the blue and green color of a painting.

In the next part of our work, degree of correlation between the auditory and visual stimuli is evaluated individually using the DCCA (1D and 2D) technique. A lower value of cross correlation exponent ($γ_x$) denotes higher level of power law correlation between the two signals involved, and vice versa. **Fig. 3 and 4** represent the values of $γ_x$ for different combinations of auditory and visual stimuli respectively. It is to be noted that in **Fig.4**, before calculating the cross-correlation coefficient for the visual stimulus, we took an average of the three cross-correlation coefficients obtained from the previous analysis for the simplification of the obtained results.

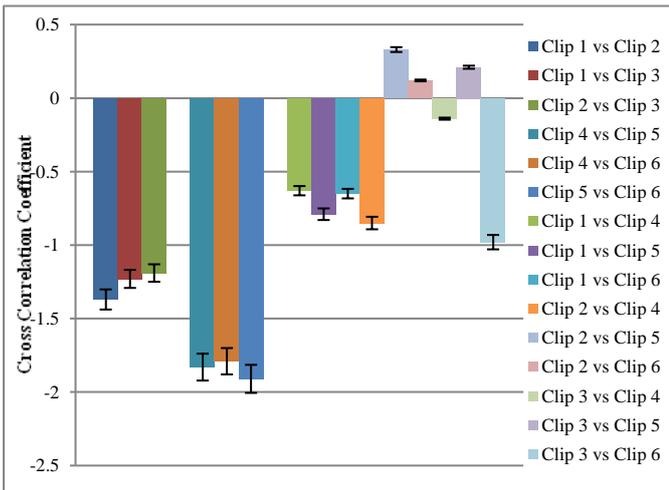 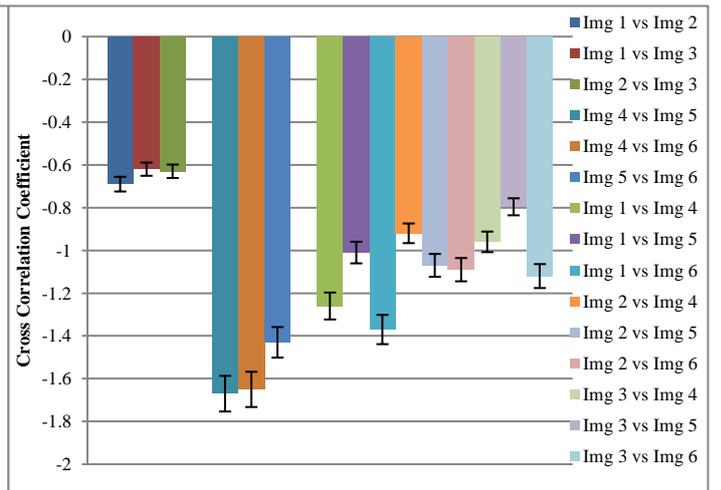

Fig. 3: Cross-correlation coefficient for different clips    Fig. 4: Cross-correlation coefficient for different images

In **Fig. 3,** it is seen that the degree of correlation for audio clips belonging to the same valence is on the higher side as compared to the clips belonging to opposite valence. The clips which have been rated as "sad" are the ones which show highest degree of correlation, while the clips rated as "happy" also show strong correlation, but lower than the "sad" ones. The inter-valence correlations are however much lower than these, while some even so "no correlation" also. **From Fig. 4,** it is seen that the degree of correlation among Images 4 to 6 are the highest of all the combinations present here, while the correlation among the Clips 1 to 3 are the lowest. Thus, we have an indirect classification of emotional appraisal even while performing DCCA also. While the images which have been rated as "sad" provide higher degree of correlation amongst them, the "happy" rated images provides lower degree of correlation. The inter-valence correlation coefficient (i.e. the degree of correlation among the happy and sad images) lie somewhere in between the two.

## CONCLUSION

In this work, we have presented a novel algorithm to automatically classify and compare emotional appraisal from cross-modal stimuli based on the amount of long range temporal correlations present in the auditory and visual stimulus put to use. The important findings of the study can be listed as under:

1. For both the auditory and visual stimulus, an averaged DFA scaling exponent of anything greater than 1.5 denotes stimulus belonging to "sad" category.
2. The DFA scaling exponent corresponding to blue and green color is the highest in case of "sad" images while the DFA exponent for "happy" images is high for red color.
3. The DCCA exponent shows that the degree of correlation is strongest among the sad clips, while the amount of correlation is lowest for the inter-valence clips.
4. The averaged degree of correlation for happy images is very low (i.e. below -0.7) while that for sad images is considerably high (i.e. greater than -1.4). The correlation between happy and sad images interestingly lies between the two (ranging between -0.7 and -1.4).
5. Pearson correlation coefficient is computed from the variation of DFA values belonging to the stimulus from two modalities. the values of which are found to be as follows:

| Happy (audio) vs. Happy (Image) | Happy (audio) vs. Sad (Image) | Sad (audio) vs. Happy (image) | Sad (audio) vs. sad (image) |
|---|---|---|---|
| 0.987 | 0.29 | -0.493 | 0.96 |

Thus, an indirect quantitative correlation is obtained between the emotional appraisal of the cross-modal bias of auditory and visual stimuli for the first time.

**ACKNOWLEDGEMENT:**
SS acknowledges the JU RUSA 2.0 Post Doctoral Fellowship (**R-11/557/19**) and Acoustical Society of America (ASA) to pursue this research AB acknowledges the Department of Science and Technology (DST), Govt. of India for providing (**SR/CSRI/PDF-34/2018**) the DST CSRI Post Doctoral Fellowship to pursue this research work.